\documentclass[%
 reprint,
 amsmath,amssymb,
 aps,
]{revtex4-2}

\usepackage{graphicx}
\usepackage{dcolumn}
\usepackage{bm}

\usepackage{xcolor}

\begin{document}

\preprint{APS/123-QED}

\title{Buckling of a monolayer of plate-like particles trapped at a fluid-fluid interface}

\author{Suriya Prakash}
\author{Hugo Perrin}%

\author{Lorenzo Botto}
 \email{Corresponding author\\
 Email address: l.botto@tudelft.nl (Lorenzo Botto)}
\affiliation{%
Department of Process $\&$ Energy,  Faculty of Mechanical,
Maritime and Materials Engineering, Delft University of Technology, Delft, The Netherlands.
}%

\date{\today}

\begin{abstract}

Particles trapped at a fluid-fluid interface by capillary forces can form a monolayer that jams and buckles when subject to uni-axial compression. Here we investigate experimentally the buckling mechanics of monolayers of millimeter-sized rigid plates trapped at a planar fluid-fluid interface subject to uni-axial compression in a Langmuir trough. We quantified the buckling wavelength and the associated force on the trough barriers as a function of the degree of compression. To explain the observed buckling wavelength and forces in the two-dimensional monolayer, we  consider a simplified system composed of a linear chain of plate-like particles. The chain system enables us to build a  theoretical model which is then compared to the two-dimensional monolayer data. Both the experiments and analytical model show that the wavelength of buckling of a monolayer of plate-like particles is of the order of the particle size, a different scaling from the one reported for monolayers of spheres. A simple model of buckling surface pressure is also proposed, and an analysis of the effect of the bending rigidity resulting from a small overlap between nanosheet particles is presented. These results can be applied to the modeling of the interfacial rheology and buckling dynamics of interfacial layers of 2D nanomaterials.

\end{abstract}

\maketitle


\section{Introduction}

The buckling wavelength of monolayers of nearly spherical particles trapped at a fluid interface under compression has been studied with both realistic  particles (Lycopodium, Chemigum) \cite{vella2004elasticity}  as well as model particles (glass beads, zirconium oxide beads) \cite{jambon2017wrinkles}. In these experiments, the particles were spread at an air-water interface and the particle layer subject to uni-axial compression in a Langmuir trough. Both the buckling wavelength and the force on the barrier, proportional to the surface pressure \cite{binks2002particles}, were measured. A mathematical model that treats the monolayer as a continuous elastic sheet captured the buckling wavelength measured in these experiments. The relation between the mechanical properties of this sheet and the particle size was obtained by assuming an effective Young modulus $E \sim {\gamma} / d$,  where $\gamma$ is the surface tension of the bare fluid/fluid interface and $d$ is the nominal sphere diameter \cite{vella2004elasticity}. According to this model, and in agreement with the experimental results \cite{vella2004elasticity,jambon2017wrinkles,pocivavsek2008stress}, the buckling wavelength of the monolayer  scales as $\sim \sqrt{\ell_c d} $, where $\ell_c = \sqrt{\gamma/(\delta \rho g)}$ is the capillary length, $\delta \rho$ is the density difference between the two fluids across the interface and $g$ is the acceleration of gravity.

Similar compression experiments on buckling of a monolayer of 2D nanomaterial particles of graphene oxide show a buckling wavelength in the range of 4 - 20 particle lengths \cite{imperiali2012interfacial}. The theory developed for monolayers of spheres overestimates the wavelength observed for  graphene oxide monolayers by at least one order of magnitude. Given the large aspect ratio of graphene oxide sheets, applying models for spheres is questionable. Therefore, a  mathematical model describing the mechanics of interfacial monolayers of plate-like particles is necessary. Developing one such model starting from data obtained with realistic nanoparticles, which is affected by variables that are difficult to control, such as polydispersity in size \cite{ogilvie2019size} and the possibility of particle-particle overlap \cite{cote2010tunable}, is challenging. With a model experimental system, in which macroscopic particles of controlled shapes are used, one can study the buckling phenomenon and associated interfacial mechanics without the complications of an actual nanoparticle system.  
 
In this paper, we study experimentally the uni-axial compression of a monolayer of millimeter-sized plate-like particles trapped at a fluid-fluid interface by capillary forces. We start with observations of a two-dimensional monolayer of hexagonal particles at an air-water interface. We then consider a linear chain of square plates (1D system). We develop a theory to explain the linear chain system which is then applied to the two-dimensional particle monolayer. In our experiments, the particles are not overlapping for most of the monolayer deformation. However, we use the one-dimensional mathematical model to discuss possible implications of small overlaps between the particles in terms of an increased effective bending rigidity of the particle layer.

In our experiments, the Bond number based on the weight of the particles is small \cite{botto2012capillary}, so the downward distortion of the fluid interface owing to the weight of the particle (minus buoyancy) is relatively unimportant. However, as we will see, when in contact the particles can displace fluid by rotating around an axis parallel to the fluid interface. This results in a gravitational contribution to the interfacial mechanics. In the linear chain case, we are able to investigate the regime in which capillary forces are dominant over gravitational forces by density matching of the upper and lower fluids.

The motivation for the current work is to better understand the compression of two-dimensional nanomaterials at fluid-fluid interfaces. Two-dimensional nanomaterials, of which the most discussed are graphene and graphene oxide, can take the form of a colloidal dispersion of nanometrically thin plate-like particles of large aspect ratios \cite{nicolosi2013liquid, vis2015water}. Recently, the use of fluid interfaces has emerged as a way to control the assembly of these systems \cite{fahimi2013density, woltornist2013conductive}. In the Langmuir-Blodgett technique, for example, a monolayer of 2D nanomaterials is adsorbed at a flat fluid-fluid interface, and the monolayer compressed by barriers \cite{cote2009langmuir}. The monolayer is then transferred to a solid substrate \cite{biswas2009novel,neilson2020tiled}.   Critical to the performance of the resulting particle coating is predicting the particle coverage in the fluid interface upon uni-axial compression in the trough, and whether the particle monolayer displays a solid-like behavior. If the particles jam at the fluid interface, the particle monolayer can buckle and the signature of this buckling is visible in the profile of surface pressure vs. barrier displacement \cite{jambon2017wrinkles, silverberg2017wrinkling,silverberg2017controlling,cote2010tunable}. The analysis of the relation between buckling wavelength and associated force on the barrier discussed in the current paper is relevant for interpreting interfacial rheology measurements with 2D nanomaterials.   
More broadly, the current investigation is carried out in the context of understanding the mechanics of particle rafts and armored bubbles and droplets, a research field that has received increasing attention recently from the soft matter physics, colloidal science and fluid mechanics communities \cite{protiere2023particle}.

\begin{figure}
\includegraphics{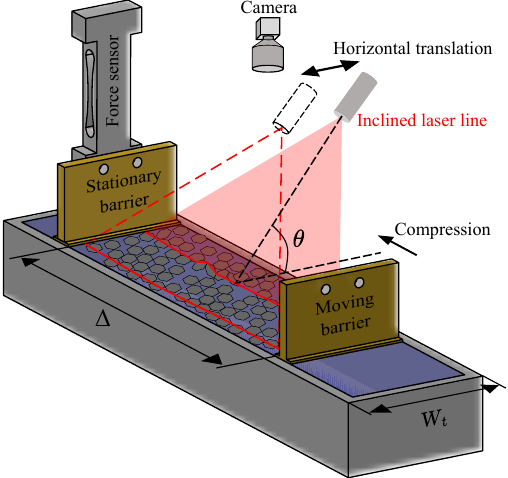}
\caption{\label{fig:setup} Schematic of the experimental setup to measure surface pressure and topology of the interfacial monolayer of plates.}
\end{figure}

\section{Experimental methods}

Uni-axial compression experiments  are carried out in an in-house-made rectangular trough of length $200\;\rm{mm}$ and width $W_t =50\;\rm{mm}$, see Fig. \ref{fig:setup}. A stationary barrier mounted on a force sensor allows us to measure the force $F$ on the barrier and the surface pressure $\Pi=F/W_t$. A moving barrier mounted on a linear stage allows us to control the distance $\Delta$ between the barriers in steps of $10\rm{\mu m}$. To measure forces of the order of $\rm{mN}$ we used a load cell with a resolution of $\pm \, 0.1\, \rm{mN}$. For small forces of the order of $\rm{\mu N}$, produced by the smallest particles we considered, we used a cantilever-based force sensor, that is described later in detail.

For the 2D monolayer experiments, we used transparent hexagonal plates made of Mylar (density $\rho_p \simeq 1400 \, \rm{kg/m^3}$) purchased from \emph{Geotech International}. The plates have thickness $t = 50\, \rm{\mu m}$ and two different lateral sizes, $L = 1.5 \, \rm{mm}$ and $3\, \rm{mm}$. Here $L$ refers to the inscribed circle diameter of the hexagonal plates. To remove possible contaminants, we aspirate the fluid interface using a suction pipette after moving the barriers to minimum opening \cite{razavi2015collapse}. The process is repeated until the fluid interface is clean. The interface is assumed to be clean if the surface pressure at maximum compression is below $4 \, \rm{mN/m}$. The 2D monolayer is prepared by gently sprinkling the particles on the air/water interface at maximum $\Delta \simeq 3 W_t$. Overlapping particles were separated by a stirring rod. The 2D monolayer is then compressed at a velocity of $200\, \rm{\mu m/s}$. The monolayer undergoes out-of-plane deformations, whose amplitude $A$ is measured by the inclined laser line method \cite{boyer2011dense,perrin2021nonlocal}. The technique involves projecting a laser sheet at an angle $\theta$ with respect to the particle-laden fluid interface (Fig. \ref{fig:setup}). The intersection of the laser sheet with the monolayer results in a  line that is imaged from the top by a camera. The intersecting line is straight for a flat monolayer and distorted for a deformed monolayer. The out-of-plane deformation amplitude can be calculated from the lateral distortion of the laser line, accounting for a proportionality constant $ \tan(\theta)$, where in our case $\theta \simeq 28^{\circ}$. The resolution of the out-of-plane deformation is $60\, \rm{\mu m}$. The novelty of our method is that we use a laser line that sweeps the monolayer providing a continuous topographic map, instead of the height profile along a single line. To do so, the laser source is mounted on a linear stage controlled by a stepper motor.

\begin{figure*}
\includegraphics{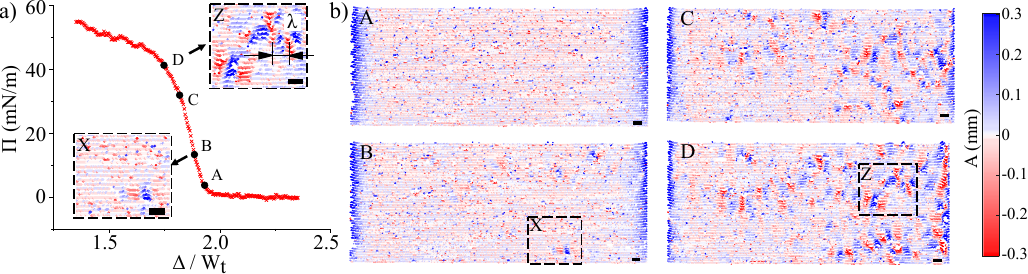}
\caption{\label{fig:2dscansforce} a) Surface pressure $\Pi$ versus normalized separation distance $\Delta/W_t$ between the barriers  for uni-axial compression of $N \simeq 1590$ hexagonal plates of lateral size $1.5\rm{mm}$ trapped at a water/air interface. We identified four characteristic points A, B, C, and D in correspondence to which the topology of the particle-laden interface is measured. The insets show zoomed-in micro-structures highlighting key features of topology in points B and D.  b) Surface topology map of the entire trough area corresponding to the characteristic points in Fig. 2 a. The blue color at the left and right ends of the topology maps corresponds to the curved menisci near the barriers. The scale bar shown as a black line is 3 mm, i.e. 2 particle diameters. }
\end{figure*}

For the single chain experiments, we used square-shaped Mylar plates of lateral sizes $L =$ 1, 3, 5, 7, 10, 15, $\&$ 20 mm and thickness $t=125\;\rm{\mu m}$, except for the $1 \; \rm{mm}$ Mylar plates for which the thickness is $t=23\;\rm{\mu m}$. For all the particles the aspect ratio $L/t$ is larger than $23$. The smallest plates are manufactured by laser cutting (\emph{Optec Laser Systems}).  Using the length and thickness of the plates and Young's modulus $\simeq 3 \ \rm{GPa}$ of Mylar, we estimate an Euler buckling threshold for the plates of $\geq 240 \ \rm{mN}$. Therefore, the plates do not buckle under compression forces of the order of a few $\rm{mN}$ and are considered to be rigid in our experiments. Experiments are carried out with both a glycerol/air interface and a water/sunflower oil interface. Corresponding density differences are $\delta \rho = 1200 \pm 1 \rm{kg/m^3}$ and $ 80 \pm 1 \rm{kg/m^3}$, respectively, measured by an Anton Paar density meter (DMA 5000). The surface tensions of the glycerol/air and water/sunflower interfaces are $65 \pm 1 \rm{mN/m}$ and $26 \pm 1 \rm{mN/m}$, respectively, measured by the pendant drop method in a Dataphysics Goniometer (OCA 25).

For the water/oil interface, the particles are first arranged at an air/water interface and the oil is gently added. Care is taken to arrange the particles in a straight chain between the barriers. Upon compression, the chain undergoes out of the plane deformation. A camera captures the side view of the chain and from the images we extracted the average amplitude $\left <A \right >$ of individual plates in the chain. As mentioned earlier, for forces of the order of $\rm{mN}$ the load cell is used. For forces of the order of few $ \rm{\mu N}$ we used a cantilever force sensor similar to the micropipette force sensor described in Ref. \cite{backholm2019micropipette}.  The deflection $\xi$ of the cantilever is measured from the side view by a calibrated camera with a zoom lens. The force is computed from $F = k \ \xi$. The stiffness $k$ of the cantilever was obtained by calibration; see Appendix \ref{app:force_sensor} for the calibration procedure and calibration curves. We used cantilevers of stiffnesses $k=29$ and $58$ $\rm{\mu N/mm}$. The resolution of the force $F$ is $\sim 1\, \rm{\mu N}$. This value is set by the  resolution of the camera ($\simeq 11 \rm{\mu m/pixel}$) and the stiffness of the cantilever.

\section{Results}
\subsection{Observations on the 2D monolayers}
\label{sec:3}
Figure \ref{fig:2dscansforce} (a) shows a typical evolution of the surface pressure $\Pi= F/W_t$ for decreasing values of the normalized distance $\Delta / W_t$ between the barriers. Fig. \ref{fig:2dscansforce} (b) shows amplitude maps corresponding to 4 characteristic points of the $\Pi \, vs. \, \Delta$ curve, denoted $A$, $B$, $C$ and $D$. For $\Delta/W_t > 2$ the plates are not touching each other and $\Pi\simeq0$, as expected. As $\Delta / W_t$ decreases, contacts between the particles are established and a non-zero value of  $\Pi$ is measured. In correspondence to point $A$, $\Pi>0$ because of the formation of force chains, but the interface remains flat (see panel A in Fig. \ref{fig:2dscansforce}). Buckling of the monolayer becomes measurable in correspondence to the point $B$. Buckling is evident from the change in amplitude of the particle-laden interface (inset X of Fig. \ref{fig:2dscansforce} (a) and inset X in panel B of Fig. \ref{fig:2dscansforce} (b)). Further compression leads to an increase in the number of  buckled regions as the surface pressure rises. The characteristic point $C$ belongs to this region of behavior. Buckling is predominantly present near the moving barrier (on the right in panel $C$ of Fig. \ref{fig:2dscansforce} b). Beyond the point $D$, referred to as   the ``collapse point'' in the following,  particle multilayers form. From $A$ to $D$, the surface pressure increases relatively steeply, while for  $\Delta/W_t$ smaller than the one corresponding to the ``collapse point'' D the surface pressure increases comparatively mildly. 

A key observation is that the characteristic wavelength of the monolayer deformation in the regions where buckling occurs is of the order of the particle diameter (see inset Z of Fig. \ref{fig:2dscansforce} (a) ). While this is shown for the $1.5\rm{mm}$ plates, the same observation holds for the larger $3\rm{mm}$ plates. Also, the monolayer does not show long-range {ordered} wave-like patterns, as reported instead for spheres \cite{jambon2017wrinkles}.  The fact that no wavelengths much larger than the particle size occur is compatible with a simple model of chain compression, which we now describe.

\subsection{One-dimensional chain model and comparison with experiment}

\label{sec:4}

\begin{figure}[h]
\includegraphics{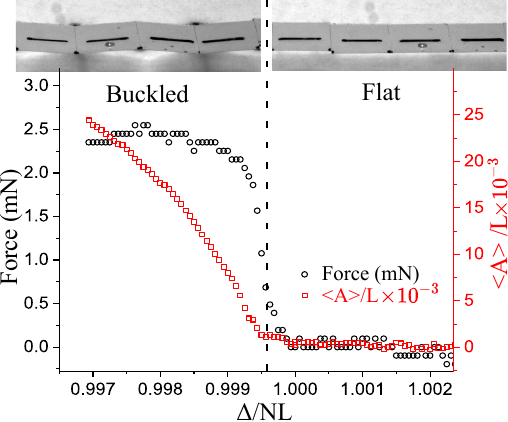}
\caption{\label{fig:1D_regimes}  Single chain compression experiment. Force $F$ (black markers) on the barrier and normalized average amplitude of out-of-plane deformation $\left< A \right > /L$ ( red markers) plotted against normalized distance $\Delta / (NL)$ between the barriers for $N=16$ square plates of size $L=10\;\rm{mm}$ at a glycerol/air interface. The vertical dashed line at $\Delta / (NL) = 0.9995$ marks the transition from the flat to the buckled state. The two insets illustrate a configuration in the flat state $\Delta / (NL) > 0.9995$ and in the buckled state $\Delta / (NL) < 0.9995$. } 
\end{figure}

We now analyze the compression of a linear chain of  $N=16$ square plates {of size $L=10\;\rm{mm}$} trapped at an air-glycerol interface.  The measured force $F$ and the {normalized} average amplitude $\left< A \right > /L$ of the out-of-plane deformation are shown in Fig.  \ref{fig:1D_regimes} as a function of $\Delta/(NL)$. From this plot, two regimes can be identified. For $\Delta/(NL) > 1$, the distance between the barriers is larger than the total length of the chain. Therefore, $F=0$ and $\left< A \right > \simeq 0$ (``flat state''). For $\Delta/(NL) = 1$, the plates touch each other and $F$ starts to increase. The measured average amplitude increases when  $ \Delta/(NL) $ is approximately equal to $0.9995$. The fact that  $F$ can be finite while   $\left< A \right > \simeq 0$, a feature that was also observed in the 2D system,  is due to small particle rearrangements before jamming. The ``buckled state'' for $\Delta/(NL) < 0.9995$ is  characterized by a sharp increase in $F$ followed by a plateau.  In the rest of this paper, we  will call the plateau value of $F$ the buckling force, as it represents the magnitude of the force that would be required to buckle the monolayer in an experiment conducted at applied force. The wavelength $\lambda$ of the monolayer corrugation was obtained by visual inspection.  Experiments with different numbers of plates, from $5$ to $16$, consistently gave $\lambda \simeq 2L$, as shown for $N=16$ in the inset of Fig. \ref{fig:1D_regimes}.

\begin{figure}
\includegraphics{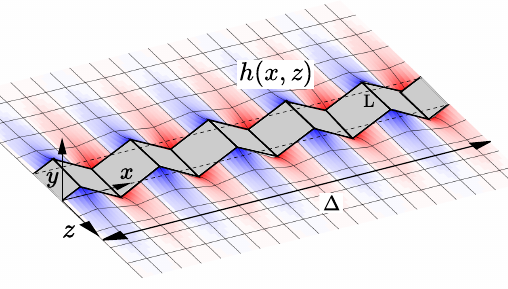}
\caption{\label{fig:interface_decay}Sketch and notations of a chain of $N$ plates of length $L$ displaced by $NL-\Delta$. The air/liquid or the liquid/liquid side menisci pinned to the edges of the particles is indicated by $h(x,z)$. The color code indicates the vertical (along the $y$ axis) deformation of the interface.}
\end{figure}

To analyze the observed behavior, we developed a mathematical model based on a balance between capillary forces, gravity and contact forces. {The vertical interface deformation caused by the plate weight is proportional to $Bo_p \ell_c$ where $Bo_p = \rho_p g L t/\gamma $ is the particle Bond number and $\ell_c = \sqrt{\gamma/(\delta \rho g)}$ is the capillary length \cite{hesla2004maximum,singh2005fluid}. From this estimate, the maximum vertical deformation is smaller than approximately $0.1 \ell_c$ for all the plates we used in our experiments. Therefore, the effect of particle weight on the interfacial distortion can  be neglected.} The total free energy of the system is then given by the gravitation potential energy of the fluid (located both below the fluid interface and below the plates), and the interfacial energy of the fluid-fluid interface. Calling $h(x,z)$ the height of the fluid-fluid interface (see Fig. \ref{fig:interface_decay}), and assuming that the plates pin the contact line at their edges \cite{yao2015capillary}, the gravitational potential energy contribution to the total free energy is 
\begin{equation}
    E_g = \int_{0}^{\Delta} dx \left [ \frac{1}{2} \delta \rho g L h^2(x,0) + 2 \int_{0}^{\infty} \frac{1}{2} \delta \rho g h^2 dz  \right] \text{,}
    \label{eqn:gravitationalpotential}
\end{equation}
where $\delta \rho = \rho_l - \rho_a$ is the difference in density between the heavier  fluid and the lighter fluids, $x$ is the coordinate along the chain and $z$ is the coordinate perpendicular to the chain in the plane of the unperturbed fluid interface, with $z=0$ corresponding to the contact line on one side of each plate (see Fig. \ref{fig:interface_decay}).
The first term in Eqn. \eqref{eqn:gravitationalpotential} is the gravitational energy of the liquid below the plates and the second term is the gravitational energy of the liquid in the two side menisci. The capillary energy associated with the  menisci on both sides of the chain is
\begin{equation}
    E_{\gamma} = 2\gamma \int_{ 0}^{\Delta} dx \left[  \int_{0}^{\infty}  \sqrt{1 + \left(\frac{\partial h}{\partial x}\right)^2 + \left(\frac{\partial h}{\partial z}\right)^2} \  dz \right].
\end{equation}
Note that we neglected the capillary contribution due to the fluid interface in the gap between the particles  (i.e. in $ -L<z<0$). To enforce the constraint that the total length of the chain is constant, we add to the total free energy the term 
\begin{equation}
    E_c = F \left[ N L - \int_{0}^{ \Delta} \sqrt{1 + \left.\left(\frac{\partial h}{\partial x}\right)^2\right|_{z=0}} \ \  \right],
\end{equation}
where $F$ is a scalar Lagrange multiplier. Physically, $F$ represents the contact force between the plates. 

\begin{figure}
\includegraphics{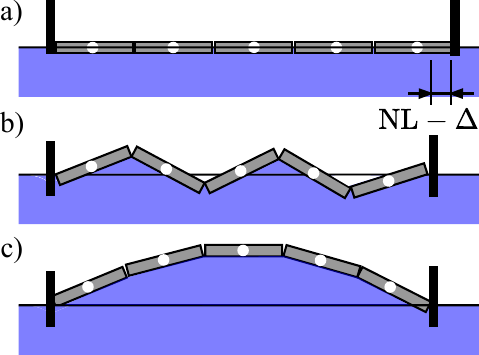}
\caption{\label{fig:different_modes} Upon compression of an initially flat monolayer (a), both configuration (b) and configuration (c) are local energy minima. We only observe configuration (b).}
\end{figure}

Setting the functional derivative $\delta \left ( E_g + E_{\gamma} + E_c \right ) = 0 $ yields two  equations. The first equation is the small-amplitude Young-Laplace equation governing the shape of the fluid-fluid interface for $-L> z > 0$:
\begin{equation} \label{eqn:laplace}
\delta \rho g h = \gamma \left( \frac{\partial^2 h}{\partial x^2} + \frac{\partial^2 h}{\partial z^2} \right).
\end{equation}
 The second equation is the boundary condition  at  $z=0$:
\begin{equation}\label{eqn:govern}
    \delta \rho g L h -2 \gamma \frac{\partial h}{\partial z} + F  \frac{\partial^2 h}{\partial x^2}  = 0. 
\end{equation}
Upon multiplication by $L$,  equation \eqref{eqn:govern} is a balance of moments. The first term represents the moment of the hydrostatic pressure force due to the weight of the fluid below the plates. The second term  represents the moment of the vertical projection of the  surface tension force at the contact line, located at $z=0$ and $z = -L$. The third term represents the moment of the contact forces $F$ between the particles.

The leading-order Fourier mode solution of Eqn. \eqref{eqn:laplace} that matches the triangle-wave profile of the contact line is \cite{lucassen1992capillary}
\begin{equation} 
    h(x,z)  = A e^{-z\sqrt{\left(\frac{2\pi}{\lambda}\right)^2 + \frac{1}{\ell_c^2}}} \sin \left( \frac{2\pi x}{\lambda}\right),
    \label{eqn:interface_shape1}
\end{equation}
where $\ell_c = \sqrt{\gamma/\delta \rho g}$ is the capillary length. Equation \eqref{eqn:interface_shape1} satisfies $h(x,z=0)=A \sin (2 \pi x/\lambda)$ and $h(x,z \to \infty) = 0$. For $\lambda \gg \ell_c$ and $\lambda \ll \ell_c$, the decay lengths of the meniscus in the $z$ direction are $\ell_c$ and $\lambda/2 \pi$, respectively. Thus, in the surface tension-dominated regime the buckling wavelength and the decay length of the fluid interface distortion are roughly of the same order of magnitude. 

\begin{figure}
\includegraphics{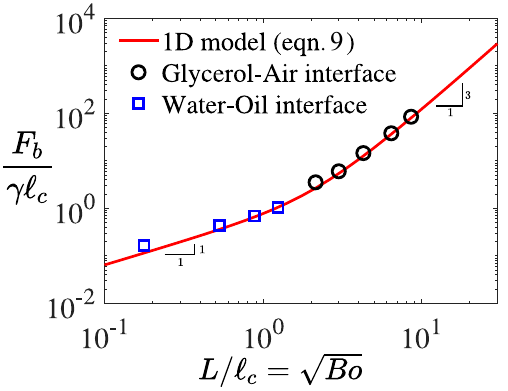}
\caption{\label{fig:1d_force} Normalized buckling force $F_b/(\gamma \ell_c)$ versus normalized particle length $L/\ell_c$ for the 1D chain experiments. The markers correspond to experimental data, blue squares  for chains at a water-oil interface and black circles for chains at a glycerol/air interface. The red line is Eqn.\eqref{eqn:force}.}
\end{figure}

Substituting \eqref{eqn:interface_shape1} into \eqref{eqn:govern} yields the contact force as a function of the wavelength:
\begin{equation}
  F =   \frac{1}{4 \pi^2} \delta\rho g L \lambda^2 + \frac{1}{2 \pi^2} \gamma  \lambda  \sqrt{(2\pi)^2 + \left( \frac{\lambda}{\ell_c} \right)^2}.  \label{eqn:force1}
\end{equation}
In Fig. \ref{fig:different_modes} we show two configurations of buckled chains, with $\lambda = 2 L$ in configuration $(b)$ and $\lambda = 10 L$ in configuration $(c)$ . Both wavelengths are local minima of the function $F(\lambda)$. The absolute minimum of $F(\lambda)$ is the total energy minimum, similar to the buckling of an Euler beam \cite{timoshenko2009theory}. Since $F(\lambda)$ is a monotonically increasing function and wavelengths smaller than $2L$ are not possible, the equilibrium wavelength is
\begin{equation}
    \lambda = 2L.
\end{equation}
The contact force corresponding to $\lambda = 2L$ is the buckling force:
\begin{equation}
    \frac{F_b}{\gamma \ell_c} =   \frac{1}{ \pi^2} \left(\frac{L}{\ell_c}\right)^3  + \frac{2}{ \pi}   \frac{L}{\ell_c} \sqrt{1 +  \left(\frac{L}{\pi\ell_c}\right)^2} . \label{eqn:force}
\end{equation}
Figure \ref{fig:1d_force} shows $F_b/(\gamma \ell_c) \, vs. \,
\sqrt{Bo} = L/\ell_c$, comparing Eqn. \eqref{eqn:force} with the experimental data.  Here $Bo = \delta \rho g L^2/\gamma$.  The agreement between the experimental data and the theory is excellent, except for the smallest values of $Bo$ where a perfect alignment of the plates cannot be ensured. For $Bo \gg 1$  the gravitational force dominates and ${F_b} \sim \delta \rho g L^3 $. In this regime, the buckling force is of the order of  the weight of the liquid displaced by each plate as the chain deforms. For  $Bo \ll 1$,  $F_b \sim \gamma L$. In this regime, the buckling force is of the order of the  capillary force exerted by the side meniscus on each plate.  Equating the first and second terms in Eq. \eqref{eqn:force} provides a threshold $L/\ell_c {\simeq} \pi$ for the transition between the capillarity-dominated  and gravity-dominated regimes.

\begin{figure}[h]
\includegraphics[width=0.45\textwidth]{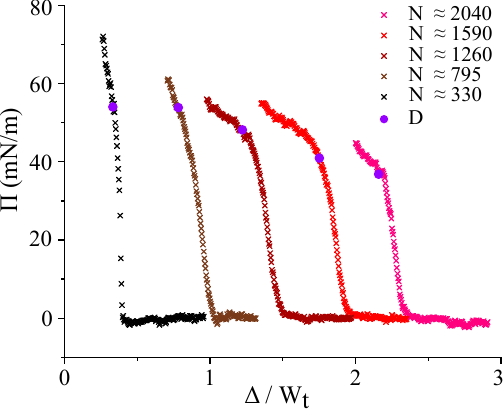}
\caption{\label{fig:2d_results} Surface pressure $\Pi$ measured at the stationary barrier for 2D monolayers of $L=1.5\rm{mm}$ hexagonal particles at a water/air interface against the distance between the barriers normalized by the trough width $\Delta/W_t$ for different number of plates. The points $D$ (in purple markers) are the collapse points.}
\end{figure}

\subsection{Comparison of 1D model with 2D experiment}
It is instructive to compare the prediction of the chain model to the experimental data for the 2D monolayer. This comparison should account for two  differences. First, in the 1D chain the internal stress in the monolayer due to particle-particle contact forces is essentially homogeneous along the compression direction (on a scale  $\gg L$). While in the 2D assembly, the contact forces are a random function of position and orientation. Secondly, in the 2D monolayer the balance of forces on the entire monolayer should account for friction with the lateral walls \cite{cicuta2009granular,saavedra2018progressive}. Evidence of the importance of the lateral walls in our experiments is the fact that the amplitude of the monolayer deformation is larger near the moving barrier (see panel C and D in Fig. \ref{fig:2dscansforce} b). A larger deformation occurs in this region because the gradient of the surface pressure along $x$ must balance the frictional stresses on the lateral walls. So the surface pressure and deformations will be larger near the moving barrier.  However, the 1D chain model could  still provide an estimate of the average value of $\Pi$ in regions where buckling occurs and sufficiently away from the lateral walls. From Eqn. \eqref{eqn:force} the buckling surface pressure is

\begin{equation} 
    \Pi_b = \frac{F_b}{L} =   \frac{1}{ \pi^2} \delta \rho g  L^2 + \frac{1}{ \pi^2} \gamma  \sqrt{(2\pi)^2 + \left( \frac{2 L}{\ell_c} \right)^2}. \label{eqn:surf_pressure_pred}
\end{equation}

\begin{figure}
\includegraphics[width=0.45\textwidth]{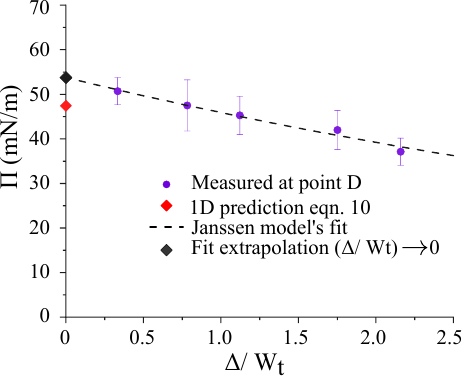}
\caption{\label{fig:2d_results_comp}Surface pressure at collapse point D for 2D monolayers of $L=1.5\rm{mm}$ hexagonal particles at a water/air interface, averaged over $3$ realizations and plotted against the distance between the barriers (normalized by the trough width). The error bars represent the standard deviations. The dashed curve is the Janssen model's fit to the experimental data. The black marker is the extrapolation of the Janssen model's fit for $\Delta/W_t \to 0$. The red marker is the surface pressure predicted by the 1D chain model Eqn. \eqref{eqn:surf_pressure_pred}. } 
\end{figure}

We performed buckling experiments at different trough aspect ratios. Trough aspect ratios are changed by varying the number of particles between the barriers for a fixed trough width. The particle sizes are fixed ($L = 1.5 \rm{mm}$) in all experiments. Figure \ref{fig:2d_results} shows the surface pressure profiles for the 2D monolayer as a function of $\Delta/W_t$ for $N \simeq 330 - 2040$. We see from this curve that the surface pressure profile depends on the initial trough area, another manifestation of the effect of lateral wall friction \cite{cicuta2009granular,  saavedra2018progressive,  andreotti2013granular}.

Figure \ref{fig:2d_results_comp} shows the experimental data for the surface pressure in the 2D monolayer, averaged over 3 different measurements, for different values of $\Delta/W_t$. This figure is obtained from Fig. \ref{fig:2d_results} by reporting the value of $\Pi$ and $\Delta/W_t$ corresponding to the collapse point $D$. In order to compare the collapse surface pressure in the 2D monolayer with the 1D model we used a Coulomb model for the lateral wall friction as done in \cite{cicuta2009granular} for a monolayer of spherical particles. This model assumes that the frictional force per unit length is proportional to the local values of $\Pi$ according to a proportionality constant $\mu_{wall}$. This approximation yields an exponential decay law also referred to as the Janssen model,  $\Pi = \Pi_0 \exp{(-2  \mu_{wall} \nu \Delta/W_t)}$. Here $\Pi_0$ is the pressure at the moving barrier and $\nu$ is the ratio of surface pressures perpendicular and parallel to the compression direction. Assuming $\nu = 1/3$ \cite{cicuta2009granular}, the best fit to the data (dashed curve in Fig. \ref{fig:2d_results_comp}) gives $\Pi_0 = 53.8 mN/m$ and $\mu_{wall} = 0.24$. For comparison, the reported friction coefficients for Mylar are in the range $0.13 - 0.41$ \cite{vella2009statics}. The black square dot in figure \ref{fig:2d_results_comp} is the extrapolation of the experimental data for the 2D monolayer to $\Delta/W_t=0$, which yields $\Pi_0 = 53.8 \rm{mN/m}$. The red square dot in figure \ref{fig:2d_results_comp} is obtained by using the parameters of our problem in Eqn. \eqref{eqn:surf_pressure_pred}. The value of $\Pi_0$ from the friction model is larger than the value from the 1D chain model, but the difference is small (about $13 \%$). Considering the simplicity of the chain model, the agreement with the 2D data is surprisingly good.

As stated before, the 2D monolayer differs from the 1D chain in the distribution of contact forces between the particles. Statistics of contact forces between jammed particles have been studied extensively in the context of granular materials \cite{PRLJasna, Majmudar2005aa, PRENagel, andreotti2013granular}. These studies reveal that the probability of contact forces attaining a value $f$ larger than the mean value $\left< f \right >$ decays fast,  approximately as $p(f/\left< f \right > )\sim\exp{(-\beta f/\left< f \right > )}$, with $\beta$ an $O(1)$ numerical coefficient  \cite{PRLJasna, Majmudar2005aa, PRENagel, andreotti2013granular}. Therefore it is expected that the monolayer contains few contact forces that are large compared to the average contact force \cite{andreotti2013granular}. Upon monolayer compression, the first buckling events will occur for groups of particles for which the contact force exceeds the estimate in Eqn.  \eqref{eqn:force}. Because such large forces are small in number, the buckling regions are initially localized, as seen in panel B in Fig. \ref{fig:2dscansforce} b. If the mechanical response of the monolayer is dominated by these spatially scattered regions, Eq. \eqref{eqn:surf_pressure_pred} could provide an upper bound for the surface pressure measured at the barrier in the 2D experiment. 

\subsection{1D model with bending rigidity}
\begin{figure}
\includegraphics[width=0.45\textwidth]{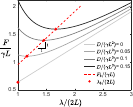}%
\caption{\label{fig:lambda_with_bending} Normalized force $F/(\gamma L)$ as a function of normalized wavelength $\lambda/(2L)$ for small values of normalized bending rigidity and $Bo = 0$ (see Eqn. \eqref{eqn:newforce}).}  
\end{figure}

Key in our derivations is the absence of bending energy in the energy functional. In an experiment with a 2D nanomaterial such as graphene oxide \cite{imperiali2012interfacial} a possible explanation for observing wavelengths larger than $2L$  could be the presence of a small but finite bending rigidity. An extension of Eqn. \ref{eqn:govern} accounting for an effective monolayer bending rigidity (per unit width) $D$ is 

\begin{equation}
     D w \frac{\partial^4 h}{\partial x^4} + \delta \rho g h w - 2 \gamma \frac{\partial h}{\partial z} + F \frac{\partial^2 h}{\partial x^2}  = 0, \label{eqn:govern2}
\end{equation}
where  $w=L$ for square particles.
Substituting Eqn. \eqref{eqn:interface_shape1} into Eqn. \eqref{eqn:govern2} gives 
\begin{eqnarray}
    \frac{F}{\gamma L} = \frac{D}{\gamma L^2}  \left( \frac{2 \pi L}{\lambda} \right)^2 \nonumber&+&2  \left( \frac{\lambda }{2 \pi L} \right) \sqrt{1 + Bo \left  ( \frac{\lambda}{ 2 \pi L} \right)^2  } \\ & &+ Bo \left( \frac{\lambda }{2 \pi L} \right)^2. \label{eqn:new_govern}
\end{eqnarray}
For $D/(\gamma L^2) \gg 1$, the buckling mechanics is dominated by competition between gravitational and bending forces. Thus,    $\lambda_b = (D/\delta \rho g)^{1/4}$, which is the result of Ref.  \cite{pocivavsek2008stress}. For $D/(\gamma L^2) \ll 1$  bending rigidity effects are negligible and we recover the results of Sec. III b. For intermediate values of $D/(\gamma L^2)$, the wavelength that minimizes the force will be larger than $2L$. Its precise value can be found by solving  $dF_b/d\lambda = 0$. For $Bo \ll 1$ the buckling wavelength is
\begin{equation}
   \frac{\lambda_b}{2L} = max \left \{1, \  \pi  \left( \frac{D}{\gamma L^2} \right)^{1/3} \right\} \label{eqn:new_wavelength}
\end{equation}
and the corresponding buckling force is

\begin{equation}
    \frac{F_b}{\gamma L} = max \left\{ \frac{2}{\pi} , \  3   \left( \frac{D}{\gamma L^2} \right)^{1/3}  \right\}. \label{eqn:newforce}
\end{equation}
Figure \ref{fig:lambda_with_bending} shows $F/(\gamma L)$ $vs.$ $\lambda/(2L)$ for $Bo = 0$ and selected small values of $D/(\gamma L^2)$. The wavelength that minimizes $F$ is indicated by the red dots. From Eqn. \eqref{eqn:new_wavelength} and \eqref{eqn:newforce} we see that both the buckling wavelength and buckling force are proportional to $(D/(\gamma L^2))^{1/3}$, thus $F_b \propto \lambda_b$ (red dashed line in Fig. \ref{fig:lambda_with_bending}). For increasing values of $D/(\gamma L^2)$ the wavelength that minimizes the force becomes larger than $2L$. 

In an interfacial monolayer of 2D nanosheets, the nanosheets can overlap slightly \cite{cote2009langmuir, goggin2021stacking, gravelle2021adsorption}. This overlap can result in a small but finite effective bending rigidity because of the  attractive force between the sheets in the overlapping region. In Ref. \cite{jones1924determination}  a Lennard-Jones potential was used to model the attractive interaction potential between parallel sheets of graphene. Using the Lennard-Jones potential, and assuming that the angle between pairs of overlapping sheets is small, it is easy to estimate the effective bending rigidity corresponding to an average overlap length $\ell$ (see Appendix \ref{app:bending_rigidity}):
\begin{equation}
    D \simeq  \frac{40 \Gamma L \ell^3}{3 r_0^2}. \label{eqn:bending_rigidity}
\end{equation}
Here $\Gamma$ is the adhesion energy per unit area and $r_0$ is the nanometric equilibrium separation between the nearly-parallel sheets. The model suggests a strong $\ell^3$ scaling with the overlap length. For  graphene oxide sheets in high-humidity conditions, molecular dynamics simulations suggest $r_0 \simeq 7.7 - 12 \ \rm{A^{\circ}}$ \cite{lerf2006hydration,dyer2015modelling} and $\Gamma \simeq 0.1 - 0.2 \ \rm{J/m^2}$ \cite{soler2018role,gravelle2021adsorption}. Taking realistic values  $\Gamma = 0.2 \ \rm{J/m^2}$,  and $r_0 = 12 \ \rm{A^{\circ}}$ and an average sheet length   $L = 1 \rm{\mu m}$, $D/(\gamma L^2)$ is estimated to be $0.02$ and $26$ for $\ell = 1 \rm{nm}$ and $10 \rm{nm}$, respectively  (assuming the surface tension of water, $\gamma = 0.07 \ \rm{J/m^2}$). The corresponding wavelengths are $2 \ \rm{\mu m}$ and $20 \ \rm{\mu m}$. Thus, even for relatively small  overlaps of only $10nm$, the wavelength of buckling can be an order of magnitude larger than $2L$.

\section{Conclusions}

We have measured the amplitude of deformation, wavelength and force on the barrier  for a two-dimensional and one-dimensional monolayer of plates trapped at a fluid-fluid interface and subject to uni-axial compression. The amplitude and wavelength of the corrugations of the 2D monolayer were measured by a laser scanning technique. 

The model we have developed to predict the experimental data for the linear chain (one-dimensional monolayer) predicts the buckling force well over a wide range of values of $L/\ell_c$, where $\ell_c$ is the capillary length and $L$ is the particle length, and  without adjustable parameters (Fig. \ref{fig:1d_force}).  The 1D chain model provides a reasonable order of magnitude estimate  of the buckling surface pressure $\Pi$ for the two-dimensional monolayer, provided that this pressure is identified as the collapse pressure corresponding to the point $D$ in Figs. \ref{fig:2dscansforce} and \ref{fig:2d_results}. The chain model does not contain a dependence on the trough aspect ratio $\Delta/W_t$, but the inclusion of frictional forces with the lateral wall via a Coulomb friction model enables us to model the observed dependence of $\Pi$ on $\Delta/W_t$.

The chain model predicts a buckling wavelength $\lambda ~=~ 2L$, independent of $L/\ell_c$. The 2D monolayer does not display  a regular  wave pattern, but the local wavelength in regions where buckling occurs is of the order of the particle size, as in the chain model.  Uni-axial compression of monolayers of spherical particles  gives smooth undulations with a wavelength $\lambda \sim \sqrt{\ell_c L}$ \cite{vella2004elasticity},  different from the one we observe. In our case, the effective bending rigidity of the monolayer is negligible, as the plates can ``hinge'' at their contact points without a bending energy penalty. In the case of spheres, even in the absence of colloidal force contribution bending energy can originate from the motion of the contact line on the surface of each particle as the mean interface curvature changes \cite{kralchevsky2005thermodynamics}. An indication of this is that the order of magnitude of the effective bending rigidity corresponding to $\lambda \sim \sqrt{\ell_c L}$ is $\gamma \ d^2$; this can be seen as the change in interfacial energy as a sphere of diameter $d$ protrudes in the fluid interface over a distance comparable to $d$.  In our case, the undulations of the contact line relative to the particles, if present, are at most limited to a scale $t \ll L$, where $t$ is the particle thickness. The corresponding changes in interfacial energy upon a change in interfacial  curvature  is $O(\gamma L t)$  \cite{yao2015capillary,cavallaro2011curvature}. For $L/\ell_c \ll 1$ and  $t/L \ll 1$, this contribution is negligible  in comparison to the dominant contribution, of order $\gamma A \lambda \sim \gamma L^2$, due to the rotation of each particle as the monolayer is compressed. The aspect ratio of the particle thus determines which capillary energy contribution controls the micromechanics of the particle monolayer.

In our experiments, we prepare the particle-laden interface ensuring no initial overlaps. If a monolayer of 2D nanosheets is prepared with care,  overlaps can be largely prevented (nanosheet stacking requires overcoming an energy barrier \cite{goggin2021stacking}), but probably not completely eliminated at large degrees of compression. Tuning the pH of the liquid \cite{cote2010tunable} or adding surfactants \cite{silverberg2017wrinkling} has been shown to suppress the stacking of 2D materials at fluid interfaces, so one may realize the experimental systems described in the current paper using real 2D materials. If particle overlaps did occur even before the compression of the particle-laden interface, the analysis would need to account for particle-particle interactions as well as statistics of the geometry of the overlapping regions. Overlaps contribute to a finite bending rigidity as a result of the adhesion forces between the nanosheets. We have shown mathematically that this effect increases the buckling wavelength compared to 2L (see Fig. \ref{fig:lambda_with_bending}).

Compression of plate-like  particles  trapped at  fluid interfaces occurs in a variety of applied settings, for instance in the manufacturing of thin films  \cite{silverberg2017wrinkling,cote2009langmuir,carey2023high},  in the deformation of Pickering emulsions \cite{kim2010graphene,vis2015water}, or in the production of crumpled graphene by aerosolization \cite{chen2012aerosol}. This work contributes to our understanding of the link between particle shape, contact mechanics, and response of the fluid interface during the compression of monolayers of plate-like particles of controlled geometry.

\section*{Acknowledgements}
We thank Simon Gravelle and Adyant Agarwal for useful discussions on modeling the interaction energy between two nanosheets. We thank Paul Grandgeorge for useful suggestions on force measurements in the $\mu N$ range.  We gratefully acknowledge funding by European Research Council (ERC) under the European Union’s Horizon 2020 Research and Innovation program (project FLEXNANOFLOW, grant agreement no. 715475).

\appendix

\section{Micro force sensor}
\label{app:force_sensor}
\begin{figure}[h]
\includegraphics{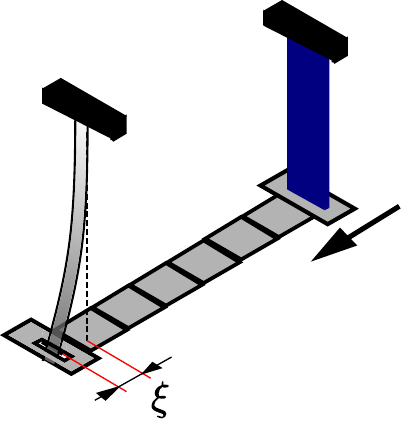}
\caption{\label{fig:force_sensor} Forces measurement with cantilever force sensor.} 
\end{figure}

The cantilever force sensors are Mylar sheets with lengths of 80 and 100 mm, a width of 10 mm, and a thickness of 125 $\mu m$. One end of the sheet is clamped and the free end is unconstrained. The free end is passed through another Mylar sheet, with a rectangular hole, which acts as the barrier (see Fig. \ref{fig:force_sensor}). The deflection of the Mylar sheet ($\xi$) from its undeformed position is calculated by imaging from the side view. 

\begin{figure}[]
\includegraphics[width=0.48\textwidth]{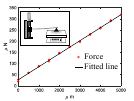}
\caption{\label{fig:sensor_calib} Calibration curve of the force sensor. The inset shows the schematic of force sensor calibration with precision balance.} 
\end{figure}

To calibrate the force sensors, the fixed end of the cantilever is mounted on a manual precision stage and the free end is rested on a knife edge placed on a Mettler Toledo precision micro-balance. Imposing successive displacements of 0.5 $\rm{mm}$ in the manual precision stage, the corresponding forces are read from the balance. Figure \ref{fig:sensor_calib} shows force $vs.$ displacement of the manual stage. The force values are linear with respect to the displacement for displacements as larger as 5 $\rm{mm}$. The slope of the line fitted to the experimental data gives the stiffness $k$ of the beam.

\begin{figure}[]
\includegraphics[width=0.48\textwidth]{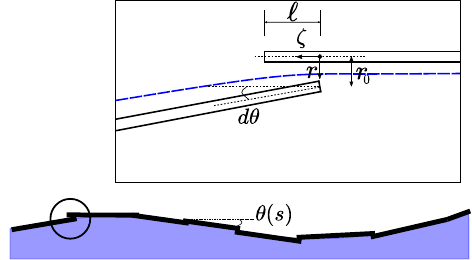}
\caption{\label{fig:overlap_sheets} Schematic of a fluid interface populated by slightly overlapping sheets. The inset shows the zoom of the overlap region. The blue dashed line is the average position of the particle-laden fluid interface.} 
\end{figure}

\section{Bending rigidity due to overlaps}
\label{app:bending_rigidity}
The equilibrium distance between two nanosheets is determined by the competition between the attractive van der Waals and the repulsive electrostatic forces between the solid surfaces. A Lennard-Jones potential has been used to model the interaction between two nanosheets in \cite{shih2010understanding, agrawal2022viscous}. We use the standard 4-10 Lennard-Jones potential energy of interaction (per unit area) between two thin parallel plates \cite{jones1924determination} 
\begin{equation*}
    \phi(r) = \frac{\Gamma}{3} \left( 5 (r_0/r)^4 - 2 (r_0/r)^{10} \right),
\end{equation*}
where $r$ is the separation distance between the plates, $r_0$ is the equilibrium separation and $\Gamma = \phi(\infty) - \phi(r_0)$ is the adhesion energy. If the separation distance $r > r_0$ the plates attract each other due to van der Waals forces and if $r < r_0$ the plates repel each other due to electrostatic forces. In the limit of small displacement around $r_0$, a quadratic approximation to the energy per unit area is \cite{agrawal2022viscous}
\begin{equation*}
    \phi(r) \simeq \frac{20 \Gamma}{r_0^2} (r - r_0)^2.
\end{equation*}
We consider a 1D chain of plate-like particles at a fluid interface where each particle pair has a small overlap of length $\ell$ (see Fig. \ref{fig:overlap_sheets}). We model the interface as a continuous curve parameterized by $\theta(s)$, the local rotation angle along the curvilinear coordinate $s$. The configuration of a single overlap is illustrated in the inset of Fig. \ref{fig:overlap_sheets}. Referring to this figure, we take $r$ in the direction normal to the top plate and $\zeta$ in the direction tangential to the top plate. Under compression the plates are rotated with respect to each other by an angle $d\theta$. The displacement of the second plate is $r (\zeta) = r_0 + \zeta \ \tan(d\theta)$ (see figure \ref{fig:overlap_sheets}). The energy required to impose this rotation for a particle pair is
\begin{align*}
    dE &
    \simeq  w\int_{0}^{\ell}\frac{20 \Gamma}{r_0^2} (r(\zeta) - r_0)^2  \ d\zeta.  
\end{align*}
Carrying out the integration for $|d \theta| \ll 1$ we obtain 
\begin{align*}
    dE &\simeq \frac{w}{2}  \left  ( \frac{40 \Gamma \ell^3}{3 r_0^2}  \right)  d\theta ^2 
\end{align*}
Multiply and divide by $(ds)^2$, where  $ds$ is an infinitesimal element of curvilinear coordinate, we obtain 

\begin{align} \label{eqn:dEexpression}
dE &\simeq \frac{w}{2}  \left  ( \frac{40 \Gamma \ell^3}{3 r_0^2} ds \right)  \left(\frac{d\theta}{ds} \right)^2 ds.
\end{align}


For a continuous surface, the bending rigidity $D$ (per unit width) is defined so that $dE = \frac{1}{2} w D \kappa^2 ds$, where $\kappa = d\theta/ds$ is the curvature. Comparing this expression to Eqn. \eqref{eqn:dEexpression} we obtain $D = (40 \Gamma \ell^3)/(3 r_0^2)ds$. In our case, because $dE$ represents the energy per particle pair, $ds$ is the distance between two particle centers, i.e.  $ds = L - \ell$. For $\ell \ll L$ the estimate of the bending rigidity is  $D = (40 \Gamma \ell^3)/(3 r_0^2)L$, as in Eq.  \eqref{eqn:bending_rigidity}. The assumption of a continuous surface is reasonable if $N \gg 1$, where $N$ is the total number of plates \cite{stockie1998simulating}. The bending rigidity thus scales proportionally to the adhesion energy $\Gamma$ and depends strongly on the overlap length $\ell$.



\bibliography{biblio}

\end{document}